\documentclass[12pt]{iopart}
\usepackage{sub_JP}
\usepackage{times}
\usepackage{amssymb}
\usepackage{enumerate}
 
\newcommand{\WW}{\mathcal{W}}

\usepackage{amsthm}
\usepackage{mathrsfs}
\let\rho=\varrho
\let\phi=\varphi
\def\real{\mathbb R}
\newtheorem{theorem}{Theorem}[section]

\newtheorem{proposition}[theorem]{Proposition}

\def\ie{{\it i.e.}}
\def\eg{{\it e.g.}}
\def\HALF{{\textstyle\frac{1}{2}}}

\newenvironment{remark}{\noindent\textbf{Remark.}}{}
\def\const{{\rm const. }}

\def\HALF{{\textstyle\frac{1}{2}}}

\usepackage[scanall]{psfrag}
\usepackage{epsfig}
\def\C#1{\begin{center}{#1}\end{center}}
\def\fref#1{Fig.~\ref{#1}}

\def\eref#1{(\ref{#1})}

\def\OO{{\cal O}}

\def\emph#1{{\it #1}}

\def\d{{\rm d}}

\newenvironment{myitem}
{\begin{itemize}
  \setlength{\itemsep}{1pt}
  \setlength{\parskip}{0pt}
  \setlength{\parsep}{0pt}}
{\end{itemize}}
\let\epsilon=\varepsilon

\def\d{{\rm d}}

\def\eref#1{(\ref{#1})}
\def\sref#1{Sect.~\ref{#1}}

\usepackage{MHequ}

\def\C#1{\begin{center}#1\end{center}}

\def\const{{\rm const.}}
\def\d{{\rm d}}
\def\dx{\d x}
\def\HALF{{\textstyle{\frac{1}{2}}}}
\let\phi=\varphi
\let\kappa=\varkappa
\def\WW{{\cal W}}

\newcommand{\W}{\mathcal{W}}

\newcommand{\fc}{f_{\rm{crit}}}

\newcommand{\fw}{f_{\rm{W}}}

\def\fref#1{Figure~\ref{#1}}
\def\sref#1{Section~\ref{#1}}
\def\eref#1{(\ref{#1})}

\def\fh{f_{\rm H}}
\def\fs{f_{\rm S}}
\def\ft{f_{\rm T}}
\def\eh{\epsilon _{\rm H}}
\def\es{\epsilon _{\rm S}}
\def\et{\epsilon _{\rm T}}

\begin{document}

\title{Noise and Topology in Driven Systems -- an Application to
  Interface Dynamics}

\author{Stewart E. Barnes${}^{1,2}$,
  Jean-Pierre Eckmann${}^{3}$, Thierry Giamarchi${}^1$, and Vivien Lecomte${}^{1,4}$}
\address{$^1$ DPMC-MaNEP, University of Geneva,  Switzerland}
\address{$^2$ Physics Department, University of Miami, Coral Gables, FL, USA}
\address{$^3$ Department of Theoretical Physics and Section de
  Math\'e\-matiques, University of Geneva, Switzerland}
\address{$^4$ Laboratoire de Probabilit\'es
  et Mod\`eles Al\'eatoires (CNRS UMR 7599), Universit\'e Paris VII, France}
\ead{$^3$ jean-pierre.eckmann@unige.ch}

\begin{abstract}
  Motivated by a stochastic differential equation describing the
  dynamics of interfaces, we study the bifurcation behavior of a more
  general
class of such equations. These equations are characterized by a
2-dimensional phase space (describing the position of the interface
and an internal degree of freedom). The noise accounts for thermal
fluctuations of such systems.

The models considered show a saddle-node bifurcation and have
furthermore homoclinic orbits, \ie, orbits leaving an unstable fixed
point and returning to it. Such systems display intermittent behavior.
The presence of noise combined with the
topology of the phase space leads to unexpected behavior as a function
of the bifurcation parameter, \ie, of the driving force of the
system. We explain this behavior using saddle point methods and
considering global topological aspects of the problem.
This then explains the
non-monotonous force-velocity dependence of certain driven interfaces.

\end{abstract}
\pacs{05.45.-a, 05.10.Gg, 75.60.Ch }
\maketitle
\section{Introduction}\label{s:intro}

In this paper, we consider a simplified, but general, description
of driven dissipative systems described by two degrees of freedom
in the presence of thermal noise.
The theory applies to systems
with two phases separated by a rigid moving domain wall (DW) with an internal
degree of freedom,
but it also describes general stochastic differential equations having a
homoclinic saddle bifurcation. We will study in detail the behavior of
such equations.

The paper is written with two audiences in mind; those interested and
familiar with dynamical systems in the presence of noise -- and those
more interested in physical applications. A short account has been given in \cite{oldstuff}.

\subsection{Physical motivation}

A large variety of physical systems have interfaces separating
different phases, with examples ranging from
magnetic~\cite{lemerle_domainwall_creep,bauer_ferre_dw-dipolar_prl2005,yamanouchi_dw_prl1996,metaxas_ferre_prl2007}
or ferroelectric~\cite{paruch_PZT_prl2005,paruch_triscone_PZT_apl2006}
domain walls, to growth
surfaces~\cite{barabasi_stanley_book,krim_growth_review}, contact
lines~\cite{moulinet_distribution_width_contact_line}.
The properties of an interface are well described at the macroscopic
level by the
competition between
(i) the elasticity, which tends to minimize the interface length, and
(ii) the local potential, whose valleys and hills deform the interface so
as to minimize its total energy.

The theory of disordered elastic
systems~\cite{kardar_phys-rep1998,giamarchi_book_young} allows one to
determine their static and dynamical features
(\eg, the roughness at equilibrium and the response to a field).
Applying a force $f$, the interface can be driven to a non-equilibrium
steady state.  A crucial feature of the zero-temperature motion is the
existence of a threshold force $\fc$ below which the system is
pinned. The system undergoes a depinning transition at $f=\fc$ and
moves with a nonzero average velocity $v$ for $f>\fc$. Close to the
transition the velocity $v\sim(f-\fc)^\beta$ is characterized by a depinning
exponent $\beta$.
In all these situations, the velocity is a \emph{monotonous} function of the force (the
more the interface is pulled, the faster it moves).
Predictions of this theory are in very good agreement with
experimental results, especially in the creep regime for interfaces in
magnetic~\cite{lemerle_domainwall_creep} or
ferroelectric~\cite{paruch_triscone_PZT_apl2006} films.

In spite of this success, there are situations where the disordered
elastic theory does not apply: for instance, one basic assumption is
that the bulk properties of the system are summarized by the
position of the interface alone.
Here, we study the case where
the position of the interface is coupled to an internal
degree of freedom and we will show how this coupling affects the
motion of the interface.
An example is provided by domain walls in thin ferromagnetic films,
where it is known that such an internal degree of freedom (a
\emph{phase}, to be detailed below) plays an important role.
\begin{figure}[h]
\begin{center}
  \includegraphics[width=.6\columnwidth]{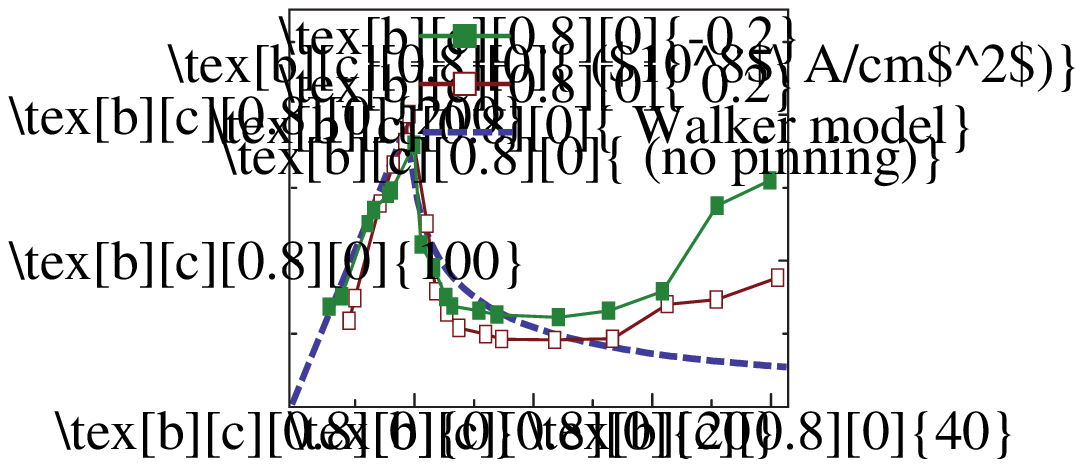} 
  \caption{The experimental velocity (dots) of an interface in a
    narrow nanowire, driven by a small external 
current, adapted from~\cite{Parkin-SCI-2008}.
  The Walker model (represented by the dashed line), which discards
  the pinning potential, does not reproduce the experimental
  results. The horizontal axis is the force (magnetic field (Oe)), the
  vertical axis is velocity (m/s).}
  \label{fig:experimental_non-monotonous_v}
  \end{center}
\end{figure}

We reproduce in Fig.\ref{fig:experimental_non-monotonous_v} 
experimental measurements of the mean velocity. It is puzzling that
the velocity is \emph{not} a
monotone function of the force.

The aim of this paper is to shed some light on this problem, by
discussing a very simplified version of the system.
We explain the general features of \fref{fig:experimental_non-monotonous_v}
by two ingredients: first, by observing that there is
a change of the topology of a typical evolution as a function of the
driving force, and second, by taking into account
temperature.
While we will work with a simplified potential, we will gain
some quite general insights on this and related problems.

The experiments mentioned above come with physical models which
describe the interaction between the phase $\phi$ and the position $r$ of the
wall. For the purposes of this paper, we will use the rigid
wall
approximation,
\cite{schryer-walker_jap-1974,malozemoff_Slonczewski_DW,slonczewski_jmmm1996,tatara-kohno_DWcurrent_prl2004,barnes_maekawa_prl2005,duine_thermalDW_prl2007}:
\begin{equa}\label{e:main1}
  \alpha \partial_t r -\partial_t \phi &= f -\cos( r)+\eta_1~,\\
  \alpha \partial_t \phi +\partial_t r &= -\HALF K_\perp \sin(2\phi)+\eta_2~.
\end{equa}
The external field $f-\cos(r)$ describes a constant ``depinning'' (or
``tilt'') force $f$ and a ``pinning'' force $-\cos(r)$ deriving from a
periodic potential. The damping coefficient $\alpha$ accounts for
Gilbert dissipation.  The effective thermal noise is a white noise with
correlations~\cite{duine_thermalDW_prl2007}
$\langle\eta_i(t)\eta_j(t')\rangle = 2(\hbar N)^{-1}\alpha k_B T
\delta(t'-t)\,\delta_{ij} $
where $N=2\lambda A/a^3$ is the number of spins in the DW, where $A$
is its cross-section, $a$ the lattice spacing.
Last, $K_\perp$ is the anisotropy constant of the ferromagnetic medium.

\subsection{Mathematical motivations}\label{s:12}

The study of \eref{e:main1} reveals that the system has a
saddle-node bifurcation at $f=\fs=1$. Furthermore, for a range of fixed
$K_\perp$
one finds values of $f$ for which the unstable manifold of the
unstable fixed point is homoclinic.\footnote{More precisely, one side
  of the unstable manifold is homoclinic, while the other goes to a
  second (stable) fixed point ($H_1$ and $S$ in
  \fref{fig:phase-portrait_f-above-fc} and \fref{fig:datacoin}).}

Saddle bifurcations have been discussed in many different contexts,
and the influence of noise is well studied. Early papers are~\cite{ETW_intermitt-noise_JPA1982}
and~\cite{bak_tang_wiesenfeld_self-organized_1988}. In those papers, the
setup is that of intermittency in the presence of noise, with a
discrete dynamical system of the form
\begin{equ}\label{e:h}
x_{i+1} = x_i -\epsilon  + x_i^2 +\xi_i +h(x_i) ~,
\end{equ}
where $x_i\in\real$, $\epsilon \in\real$ is the bifurcation parameter,
and $\xi_i$ is some appropriate noise. The term $h(x)$ describes a
function which, \eg, vanishes in the neighborhood of $x=0$ but is such
that orbits must eventually return to a neighborhood of $0$.
For this setting, under weak additional assumptions, one can study in
quite some detail the invariant measure, and several salient features
appear:
\begin{itemize}
\item Orbits stay a very long time close to $x=0$ and do fast excursions
  away from the neighborhood.
\item When the parameter $\epsilon $ is positive, the deterministic 
  system has a stable and an unstable fixed point (close to
  $x=\pm\sqrt{\epsilon }$). The stochastic dynamics then helps the
  system to escape from the attracting fixed point (which is at $\sim
  -\sqrt{\epsilon }$), but this may take a long time.
\end{itemize}

In this paper, we discuss a similar scenario, but with some new
features:
We consider the parametrization $f=1-\epsilon ^2$.
\begin{enumerate}
\item There are two equations (and they are differential equations
  rather than iterations), with a saddle-node bifurcation at the value
  of the bifurcation parameter $\epsilon =\es =0$.
\item Close to $\epsilon =0$ there is an $\eh >0$ for
  which the unstable manifold (of the unstable fixed point)
  \emph{returns} to the
  unstable fixed point.\footnote{This is called a homoclinic connection.} We will be interested first in what happens
  for $\epsilon \in (\es,\eh)$.
\item We then discuss how the topological type of the orbits can
  change when the phase space is a torus. This will lead to a
  non-monotonous mean sojourn time near the unstable fixed point.
\end{enumerate}

The normal form of \eref{e:main1} is obtained by various rescalings, and
a non-linear coordinate transformation. The deterministic part is
given by
\begin{equa}
  \dx  &=  (\epsilon x+ x^2) \,\d t  ~,\\
   \d y &=  -y\, \d t~.\\
\end{equa}

While the deterministic part
follows in a quite simple way, there is also a term appearing from the
change-of-variables (the It\^o term).\footnote{We thank a referee for
  pointing out our oversight in not discussing this term in an earlier
  version of the paper.} This term takes the form $-\sigma^2 Q \d t$
(in the $x$-component above, and a similar term for the $y$-component) with
\begin{equ}
  Q=\frac{1+\alpha ^2}{8 K_\perp \alpha ^2} +\OO(\epsilon^{1/2} )~,
\end{equ}
when $f=1-\epsilon^2 $.
We will study \eref{e:main1} in the regime where $\epsilon >0$ and
$\sigma^2\ll \epsilon $. (The simulations of \fref{fig:forcevelocity}
were done for $\epsilon> 0.03$ and $\sigma^2<8\cdot 10^{-7}$.)

Therefore, we continue the discussion of the local equation near the
fixed point with the more easily tractable form
\begin{equa}\label{ee:1}
  \dx  &=  (\epsilon x+ x^2) \,\d t + \sigma_x \d\xi_x~,\\
   \d y &=  -y\, \d t+\sigma_y \d\xi_y~,\\
\end{equa}
and omit the It\^o term.
Here, $\d\xi_x $ and  $\d\xi_y$ describe the white noise, and the three parameters are
$\epsilon \ge 0$ and $\sigma_x\ge0$, $\sigma_y\ge 0$.
Adding a term $h$ as in \eref{e:h} on can achieve that \emph{globally} the unstable manifold of $x=y=0$
returns to $x=y=0$ for some small $\eh>0$ when $\sigma=0$. We will
tacitly assume that such a term is present.
The phase space of \eref{e:main1} is the torus $(r,\phi)\in
[0,2\pi)\times[0,\pi)$ and the unstable fixed point is at
$r=0,\phi=0$. 
We will argue in Sect.~\ref{2var} that the term $\sigma_y \d\xi_y$ can
be omitted without changing the qualitative behavior of the problem.

\section{Results}
\label{sec:results}

\begin{figure}[h]
\centering
\includegraphics[width=.7\columnwidth]{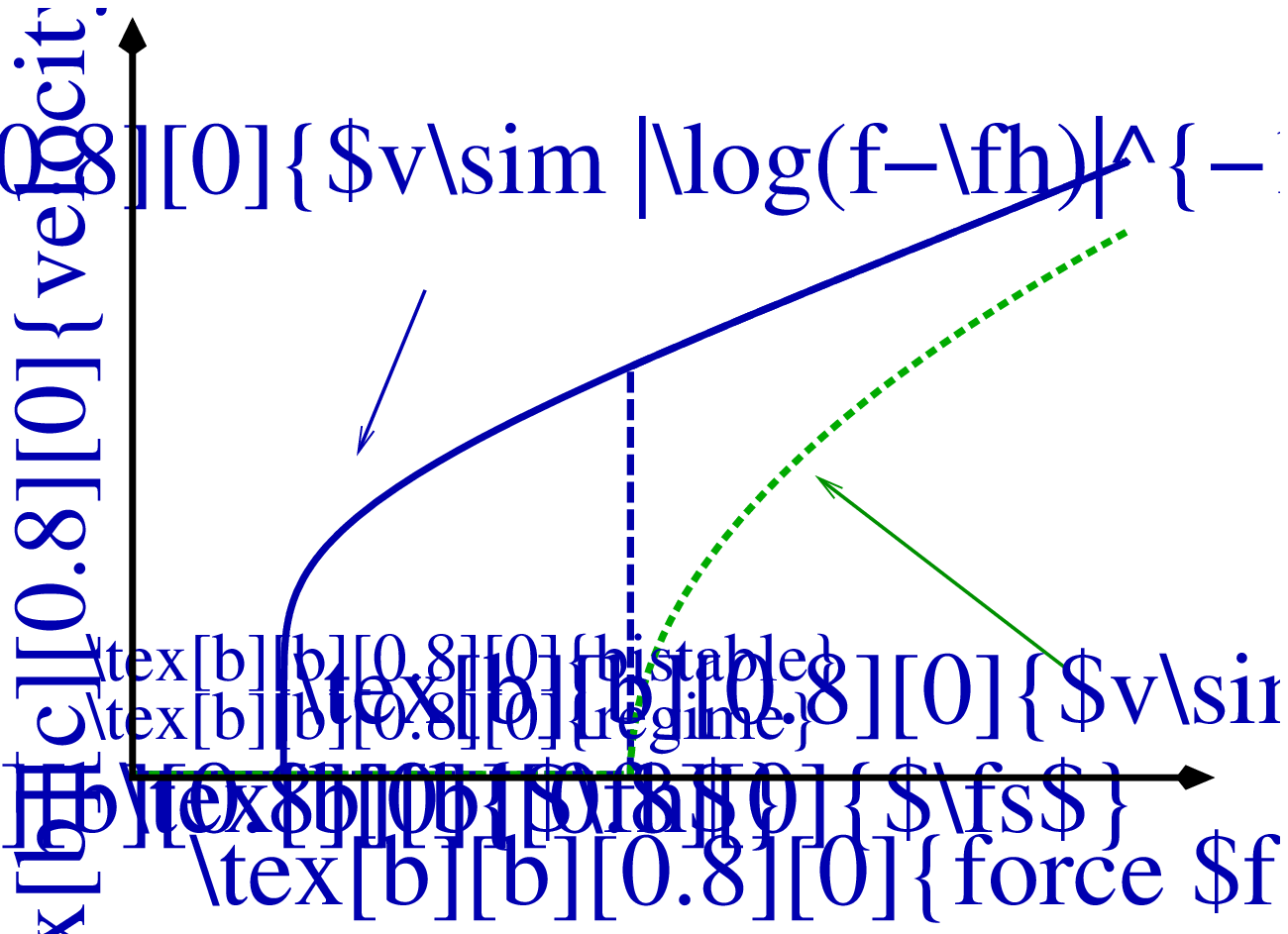}
\caption{The two scenarios for the mean velocity: The green (dashed)
  curve shows $v$ for the case of \eref{e:washboard} where no internal
  degree of freedom is present, which corresponds to the motion of a
  particle on an inclined ``washboard''. The blue (solid) curve shows
  the velocity $v$ for the case of \eref{e:main1} where the domain
  wall position $r$ is coupled to $\phi$.}
\label{fig:depinning-bistable-scheme}
\end{figure}

We first present the results from a physicist's perspective:

In \fref{fig:depinning-bistable-scheme} we illustrate the
first two findings which appear because a second field $\phi$ comes
into play: 
\emph{
\begin{myitem}
\item The critical force, at which depinning initiates, moves from
  $\fs$, ($\fs=1$), to a lower value $\fh$. Between $\fh$
  and $\fs$ the system is bistable: the velocity is either $0$ or
  strictly positive (see \fref{fig:depinning-bistable-scheme}).
\item The critical exponent of the velocity at depinning changes from
  $\HALF$ to ``$+\infty$'': the velocity grows like $v\sim 1/|
  \log(f-\fh)|$.
\end{myitem}
}
The physical picture behind the bistable regime is the following:  The
position $r$ represents the position of a particle in a tilted
periodic potential. For $f>\fs$ this potential presents no local
minima and the velocity is positive. For $f<\fs$ there are local
minima that cannot be overcome in the absence of $\phi$ (this
corresponds to the dashed curve of
\fref{fig:depinning-bistable-scheme}).  The phase $\phi$ acts as an
``energy store'' for the position $r$.  If $r$ starts close to a local
minimum, dissipation makes it end at the minimum and the steady
velocity is zero (this is the lower branch of the bistable regime in
\fref{fig:depinning-bistable-scheme}).  On the other hand, if $r$
starts far from a local minimum, the system reaches a stationary regime
where $\phi$ helps $r$ to cross the energy barriers between
successive minima, by periodically borrowing and giving ``kinetic
energy'' to $r$ (this is the upper branch of \fref{fig:depinning-bistable-scheme}).

\begin{figure}[h]
      \C{\includegraphics[width=.6\columnwidth]{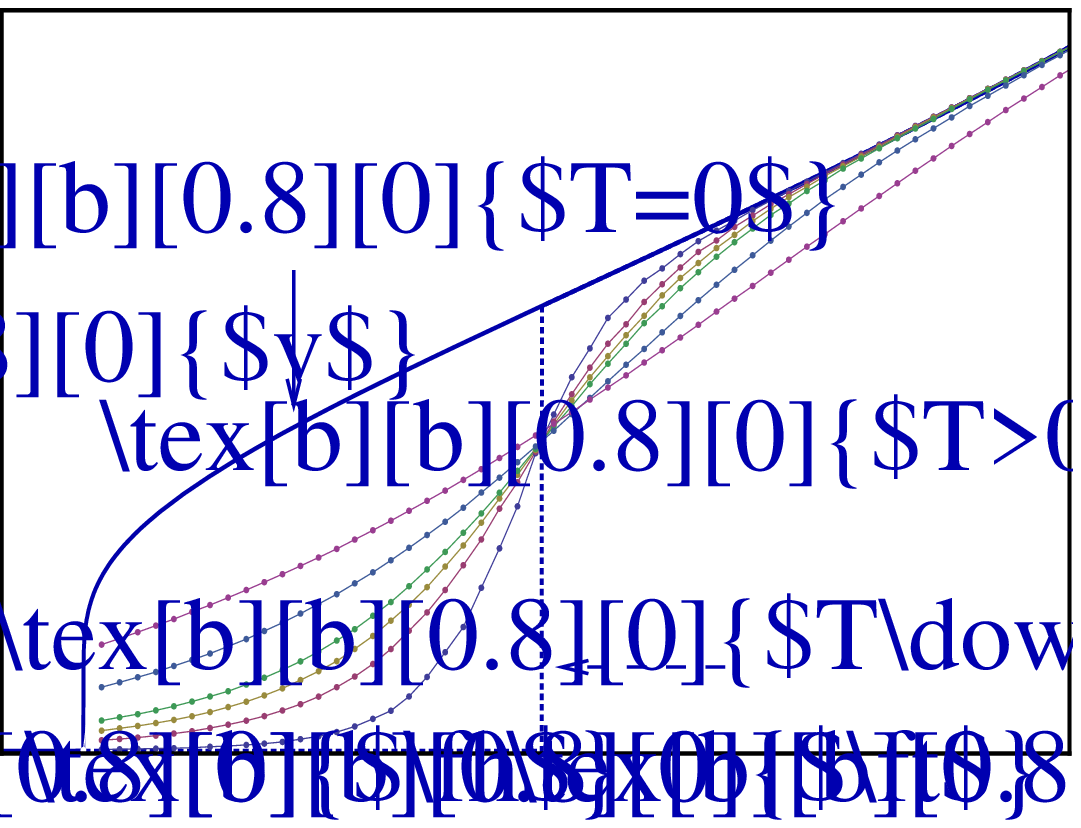}}
      \caption{The velocity for \eref{e:main1} as a function of $f$ and
      for several small values of the temperature $T$. Note that the
      curves cross at some value $\ft$. In the  limit
      $T\downarrow 0$ the curves accumulate at $\ft$, while
      the deterministic equation (\ie, $T=0$) leads to the blue
      curve. The parameters are $\epsilon^2\in\{0.03,0.1\}$,
      $\sigma\in\{0.0001.0.0009\}$, $\alpha =1/2$ and $K_\perp=6$. The
      integrator was Euler-Maruyama, with a time step of $0.0003$. We
      averaged over 512 samples.}\label{fig:forcevelocity}
\end{figure}

Furthermore, if we introduce temperature, \ie, some external noise,
then the force-velocity curve no longer presents any bistability (see
\fref{fig:forcevelocity}). This leads to a third observation:
\emph{
\begin{myitem}
\item  The appearance of a third critical force,
$\ft$. Note that for all (small) positive temperatures $T>0$,
the force-velocity curves actually cross at $f=\ft$, as illustrated in \fref{fig:forcevelocity}.
\end{myitem}
}

Our next result is a consequence of the periodicity of the r.h.s.~of
\eref{e:main1} in $\phi$. This periodicity is typical of domain walls
in magnetic systems.
The domain wall position is generically coupled to an
internal degree of freedom (for example a phase $\phi$). \footnote{Although
  this coupling is well known in the magnetic DW
  community~\cite{malozemoff_Slonczewski_DW}, it has to our knowledge
  always been discarded in interface physics.}
In \sref{sec:Global} we will show how the periodicity influences the
mean velocity
\emph{
\begin{myitem}
\item  The mean velocity of the system \eref{e:main1} is a non-monotonous
  function with many maxima (depending on the values of $\alpha $ and $K_\perp$).
\end{myitem}}

\begin{remark}
In early work \cite{schryer-walker_jap-1974},
it was observed
that, in the absence of pinning, \ie, because of the cosine in \eref{e:main1},
$v(f)$ increases up to a characteristic force
(called the Walker force) $\fw$
above which the velocity {\it decreases\/} for a large range of $f>
\fw$\, see \fref{fig:experimental_non-monotonous_v}. What we show is
that the pinning potential leads to a very different scenario.
\end{remark}

\section{A simple example}

Leaving for a moment \eref{e:main1} aside, we study first a much
easier problem to familiarize the reader with our approach.
We consider the problem of depinning from a periodic
potential, but without the phase $\phi$.
The common
underlying ingredients of such systems is the ``pulling'' of an
interface by a force $f$. As the easiest example, we can consider the
case of an ``inclined washboard'':
\begin{equ}\label{e:washboard}
\partial_t r=f -\cos( r)~,
\end{equ}
where $f$ is the constant force and $r=r(t)\in\real$ is the position of
the DW at time $t$.
Clearly, the r.h.s.~of \eref{e:washboard} can vanish only if
$|f|\le1$, and in that case every initial condition $r_0=r(0)$
will, as time evolves, converge to one of the values
$r_*=\arccos(f)+2\pi n$, with $n$ any integer. In this case, we say
that the potential is \emph{pinning}. On the other hand, when $|f|>1$,
there is no fixed point for \eref{e:washboard} and $r(t)$ will increase
or decrease indefinitely. In fact, one can check that, for
$f>1$ and $r(0)=0$
the solution of \eref{e:washboard} is:
\begin{equ}
r(t)=  2\,\arctan \left( {\frac {\tan \left( \HALF\,t\sqrt {{f}^{2}-1} \right)
\sqrt {{f}^{2}-1}}{f+1}} \right) ~,
\end{equ}
whose derivative is a periodic function of $t$ with period
$p=2\pi/\sqrt{f^2-1}$. Therefore, $r(p)-r(0)=2\pi$, and
the limit is
\begin{equ}
  \lim_{t\to\infty } \frac{r(t)}{t}=\frac{\sqrt{f^2-1}}{2\pi}=\OO(\sqrt{f-1})~.
\end{equ}
In other words, the mean displacement, which we call the \emph{velocity} $v$,
is given by $v(f)=\sqrt{f^2-1}/(2\pi)$.
Thus, for the simple case of \eref{e:washboard} the well-known result
is that near the depinning transition, the velocity grows like $\sqrt{\fs-f}$ where
$\fs=1$ in our simple example.
Furthermore, the velocity is obviously a monotone function of $f$: The
harder one pulls, the faster one advances.

\section{The coordinates of the problem at zero temperature}

We now study the special case of \eref{e:main1} when the noise
terms are absent
\begin{equa}\label{e:system}
  \alpha \partial_t r- \partial_t \phi=&~ f - V'(r) ~,\\
\alpha \partial_t \phi+\partial_t r =&  -\HALF K_\perp \sin (2\phi)~,
\end{equa}
where $V'(r)=\cos(r)$.

When $K_\perp$ is very large, $\phi$ will be very close
to $0\ {\rm mod}\, \pi$, and then the system reduces to the washboard model
\eref{e:washboard}.
However, for smaller $K_\perp$, the phase $\phi$ matters and this is
the case we want to study now.
A redefinition of $f=1-\epsilon^2 $ brings the problem \eref{e:system} to the more
convenient form
\begin{equa}\label{e:main1a}
  \alpha \partial_t r -\partial_t \phi &= -\epsilon^2  + (1-\cos( r))~,\\
  \alpha \partial_t \phi +\partial_t r &= -\HALF K_\perp \sin(2\phi)~.
\end{equa}
The phase space of this equation is the torus $(r,\phi)\in
[0,2\pi)\times[0,\pi)$.
For the following discussion the reader is
    referred to \fref{fig:phase-portrait_f-above-fc} where the torus
    is drawn in the plane with the horizontal axis corresponding to
    $r$ and the vertical corresponding to $\phi$. A three-dimensional
    rendering is shown in \fref{fig:datacoin}.
\def\bothsize{0.54}
\psfragscanoff
\begin{figure}[t]
  \centering
 \includegraphics[width=\bothsize\columnwidth]{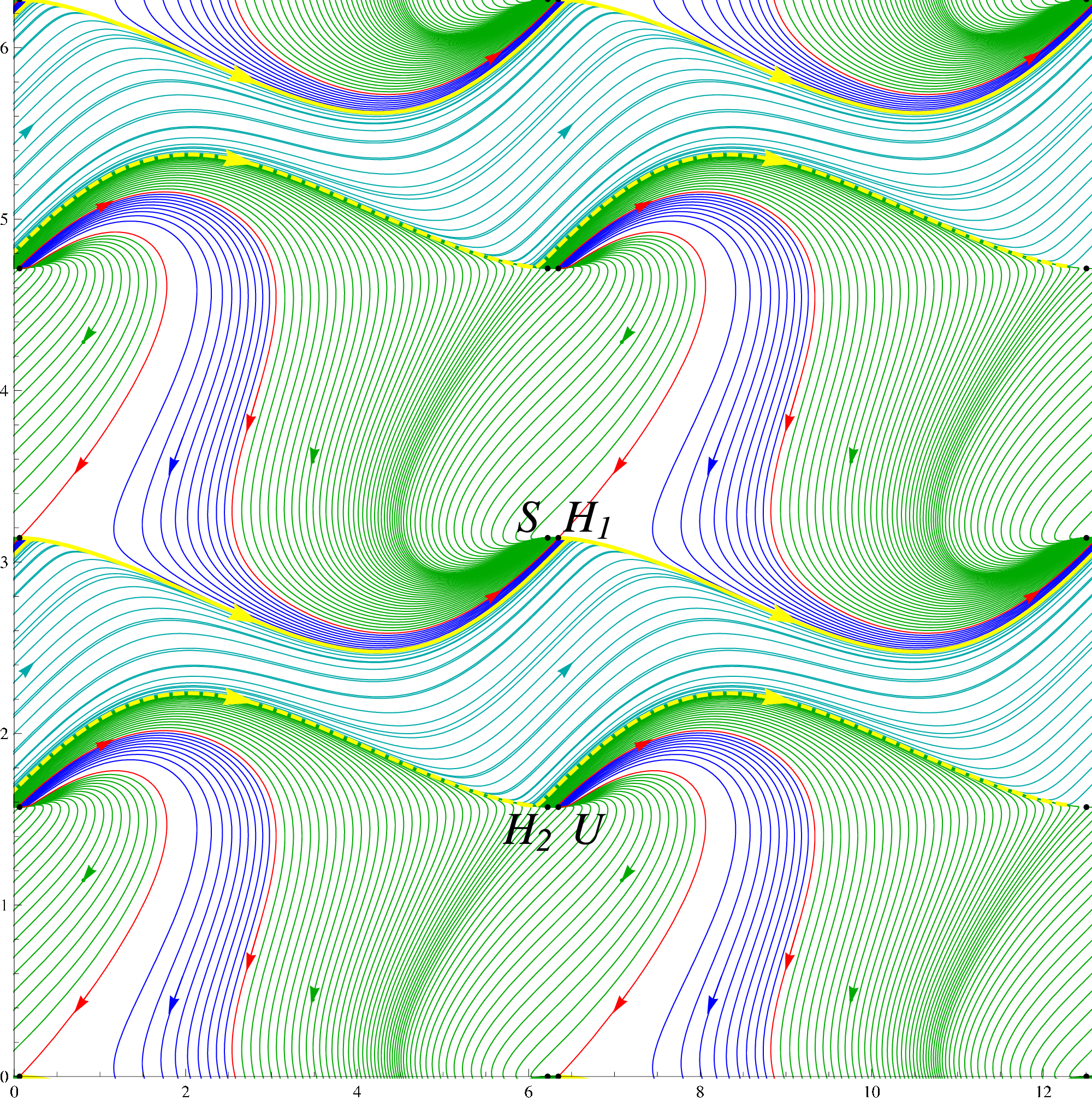}
 \caption{Phase portrait of the evolution \eref{e:main1}
 in the $(r,\phi)$ plane, for $\es=0<\epsilon<\eh$.
 %
 Boundaries between different regions are in { red}.
 The attracting limit cycle is shown as a yellow line, while the
 repulsive limit cycle is shown as a yellow dashed line.
 %
 In {green}, the basin of attraction of
 the stable fixed point $S$ ; in {
   blue}, the basin of attraction of the stable limit cycle, either
 from below (light blue) or from above (dark blue).  }
 \label{fig:phase-portrait_f-above-fc}
\end{figure}
   \begin{figure}[h]
\centering
      \includegraphics[width=\bothsize\columnwidth,angle=345]{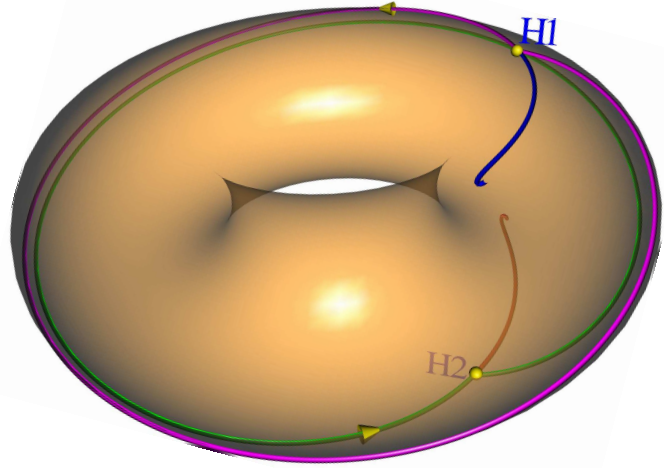}
      \caption{The unstable manifold (purple) of the
        (yellow) hyperbolic fixed point $H_1$ winds around the torus
        once (counterclockwise)
        and ends at the fixed point $H_1$. In green (behind)  the same
        for the other fixed point $H_2$. The stable fixed point is at
        the end of the blue ``tail'', and the unstable at the end of
        the orange tail.}\label{fig:datacoin}
    \end{figure}
\psfragscanon

It is easily verified that \eref{e:main1a} is invariant under the
symmetry:
$r\to -r$, $\phi\to -\phi+\pi/2$, $t\to -t$. This makes the phase
space centrally symmetric, but we will not make use of this property
in the analysis.

We will consider only values of $\epsilon^2 \ge 0$, $K_\perp>0$. For the simulations,
we took $\alpha =\HALF$.
Under these assumptions, the \emph{local} structure of this equation
is characterized by 4
fixed points of the form $(0,\pm r_\epsilon)$ and $(\pi/2,\pm r_\epsilon)$,
where
$$
r_\epsilon= \arccos(1-\epsilon^2 )~.
$$
The stability of the 4 fixed points is as follows (for $\epsilon >0$):
\begin{myitem}
\item $H_1=(0,r_\epsilon)$ and $H_2=(\pi/2,-r_\epsilon)$ are hyperbolic (with one stable and
  one unstable direction),
\item $S=(0,-r_\epsilon)$ is stable,
\item $U=(\pi/2,r_\epsilon)$ is unstable.
\end{myitem}
For $\epsilon =0$ we have $r_\epsilon=0$ and the corresponding pairs of fixed points
collide, leading to a single fixed point with one direction stable,
and the other stable-unstable. Thus, at $\epsilon =0$
the fixed points $S$ and $H_1$ (resp.~$U$ and $H_2$)
\emph{collide}; we are in the presence of a typical
\emph{saddle-node bifurcation}.

\section{General discussion for the case of non-zero temperature}
\label{sec:General}
Apart from its interest as a physics problem, the equations under study
are a nice example of the interplay of homoclinic orbits, collision of
a saddle-node, and the influence of noise. While any combination of
two of the three phenomena is amply discussed in the literature,
\cite{risken_FP}, as far as we know, the combination of all three seems to be
new. In particular, as we shall show, the system will have a ``phase
transition'' as the noise goes to zero, which occurs neither at the homoclinic point, nor at the
collision of the saddle-node, but at a well-defined intermediate
point. The present section will derive this in a general form.

\subsection{The one-variable case}
In very early work, Risken \cite{risken_FP}, considered the problem
$$
\partial_t^2 r =- \gamma \partial_t r - \epsilon  +(1-\cos(r))~.
$$
If we write it as a first order system, we have
\begin{equa}
\partial_t x &= v~,\\
\partial_t v &= -\gamma v -\epsilon  +(1-\cos(x))~.
\end{equa}
The phase space for this system is $(x,v)\in [0,2\pi)\times
  \real$. There are now only two fixed points:
$v=0$, $x=x_*\equiv \pm \arccos(1-\epsilon )$. So, the system is really quite different
  from our model. However, two of its main features
  remain and they can be discussed in the spirit of \eref{e:main1}: Locally, there are two fixed points: One is stable
  $(v,x)=(0,x_*)$ and the other $(0,-x_*)$ is hyperbolic.
  Again, for $\epsilon =0$ there is collision of the two fixed points
   (a saddle-node bifurcation). On the other hand, there is a value
  $\eh$ of $\epsilon $ (not the same as in our model) depending on $\gamma$
  for which we have a homoclinic connection).

\subsection{The 2-variable case}\label{2var}
We consider again \eref{e:main1}, but change coordinates
immediately to a normal form. Furthermore, for the purpose of the
discussion in this section,
it is irrelevant that the natural phase space is the
torus. In fact, it suffices to consider a local coordinate system near
the saddle-node. The global aspects only have to do with the ``reinjection'' \cite{ETW_intermitt-noise_JPA1982}.

In a local coordinate system where the hyperbolic fixed point $H_1$ is
at the origin, up to terms of higher order, and neglecting the It\^o
term, as discussed in \sref{s:12}, the system
can be written in the form
\begin{equa}\label{e:1}
  \dx  &=  (\epsilon x+ x^2) \,\d t + \sigma \d\xi~,\\
   \d y &=  -y\, \d t+\sigma_2\d\xi_2~.\\
\end{equa}
Here, $\d\xi $ and $\d\xi_2$ describe white noise, and the three
parameters are 
$\epsilon \ge 0$, $\sigma\ge0$, and $\sigma_2\ge0$.

We will omit
the noise term in the (stable) $y$ direction because it would only induce a
fluctuation in the ``arrival time'', but has no influence on the escape
rate to the basin of attraction of the fixed point $(-\epsilon ,0)$.
However, the noise in $x$ direction is
essential for our discussion.

\begin{figure}
\begin{center}
  \C{\includegraphics[scale=0.4]{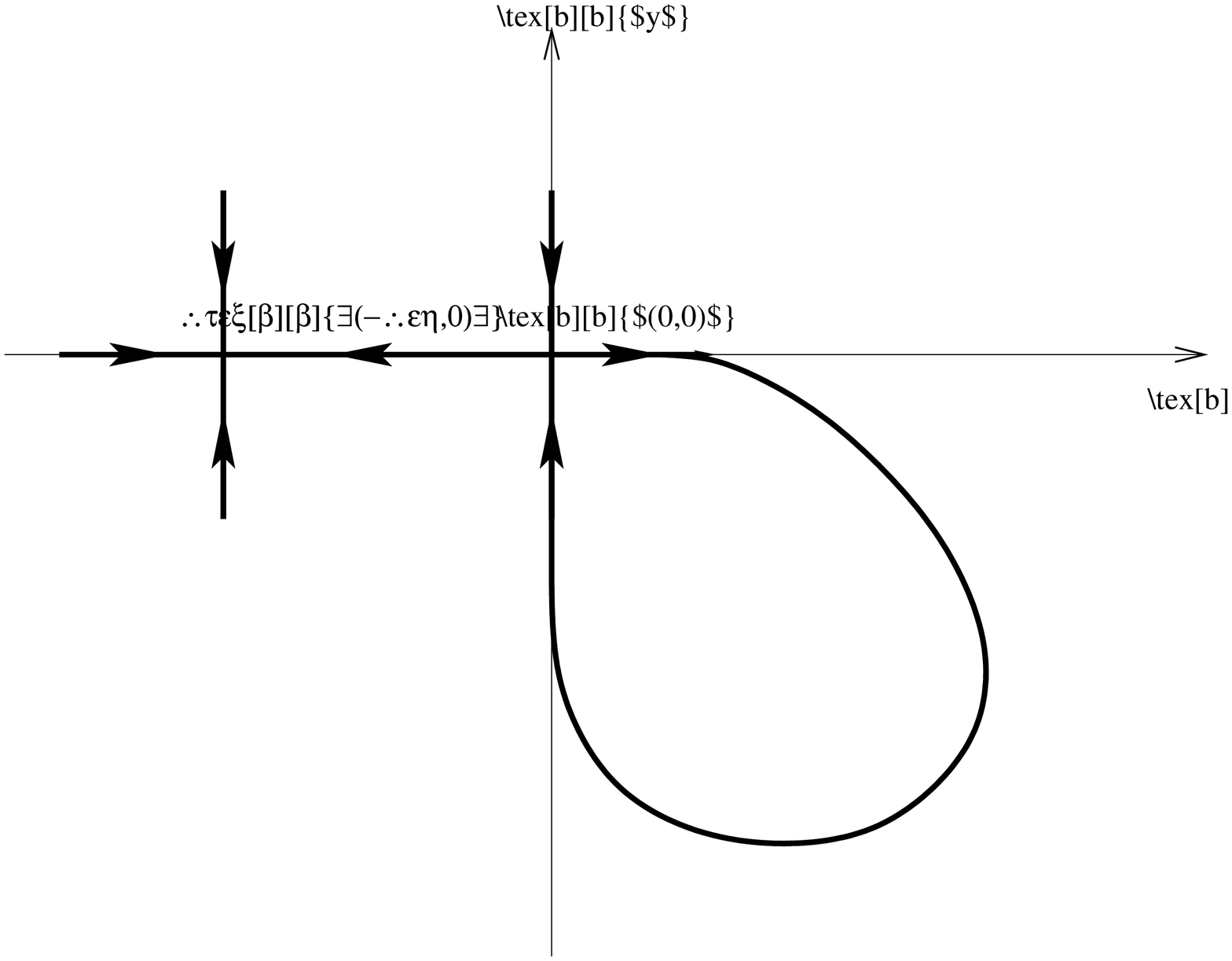} }
  \caption{The phase portrait at $\epsilon =\eh$.}\label{fig:1}
  \end{center}
\end{figure}

There is one more, crucial,
assumption:
\emph{For some $\eh>0$ (when $\sigma=0$) the unstable manifold of the fixed
  point $(x,y)=(0,0)$ (in the positive direction) is homoclinic, that
  is, it returns to $(0,0)$. Furthermore, for $\epsilon <\eh$ the
  unstable manifold is moved to the right (positive $x$).} See
\fref{fig:1}. We also assume that
this unstable manifold is transversally stable, that is, nearby orbits
are attracted to it, as illustrated in \fref{fig:1}. Such behavior can
be obtained if in \eref{e:1} we add some nonlinear terms which bend
the unstable manifold of $(0,0)$ as shown in \fref{fig:1}. \emph{We
  assume in the following that \eref{e:1} has been modified accordingly,
  without changing the vector field near $(0,0)$.}

\begin{proposition}
  Under the above assumptions, there is a constant $A>0$ such that the
  mean velocity of the system \eref{e:1} has a phase transition at a
  point $\et$ and, for small $\epsilon $ and large
  $\epsilon ^3/\sigma^2$, this transition happens at
$
\et
$
close to the solution of $$ \epsilon -6(A \cdot (\eh/\epsilon -1))^2 =0.
$$
(This solution lies in the interval $(\es=0,\eh)$.)
\end{proposition}

\begin{figure}
\begin{center}
  \C{\includegraphics[scale=0.4]{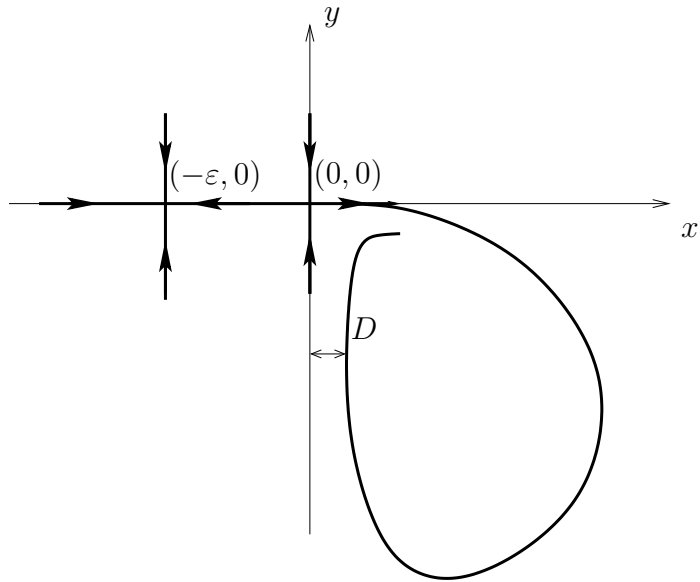} }
  \caption{The phase portrait at $\epsilon < \eh$. The
    unstable manifold of $(0,0)$ has moved to the right by a distance $D\sim
    A\cdot(\eh -\epsilon )$~.}\label{fig:2}
  \end{center}
\end{figure}

The essential thing here is that $\eh >0$. When $\sigma
>0$ the following happens: If we start at some point of the
unstable manifold, and evolve with the noisy evolution, the orbits
come back, for $\epsilon $ between $\es=0$ and $\eh$, as a
basically Gaussian distribution around the unstable manifold, see \fref{fig:2}.

\begin{figure}
\begin{center}
  \C{\includegraphics[scale=0.4]{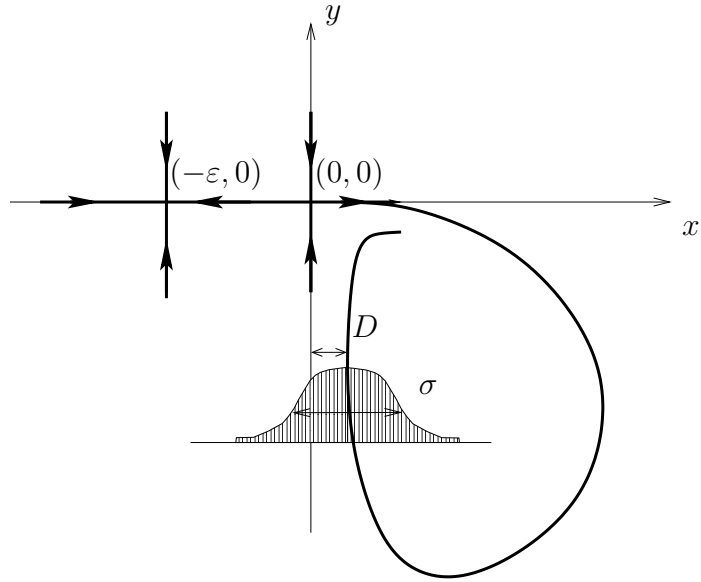} }
  \caption{The same phase portrait as in \fref{fig:2}. Superposed is
    the (Gaussian) distribution of noisy orbits returning along the
    unstable manifold of $(0,0)$. Note that the relation between $D$
    and the width $\sigma$ of the distribution determines how
    frequently a noisy orbit will fall onto the stable (left) side of
    the $y$ axis.}\label{fig:3}
  \end{center}
\end{figure}

At this point, an intriguing competition between two phenomena
occurs. On one hand, because $\sigma>0$, some orbits (those on the
``inside'' of the homoclinic loop in \fref{fig:2}) are accelerated by
the noise, since they avoid the close passage by the fixed point
$(0,0)$. On the other hand, those which return to $(0,0)$ on the side
of $x<0$ fall into the basin of attraction of the fixed point
$(-\epsilon ,0)$ and they will need a long time to escape from that
basin. This phenomenon has been studied long ago under the name of
``intermittency in the presence of noise''
\cite{ETW_intermitt-noise_JPA1982}. In that case,
it was always (rightly) assumed that the reinjection density is close
to uniform across the basin. \emph{In the case at hand,
the novel problem is
that the probability to fall into the basin of attraction
of the point $(-\epsilon ,0)$ decreases as $\epsilon $ decreases from
$\eh$.
This is because the center of the probability distribution of orbits
moves away from the basin as
  $\epsilon $ decreases, see \fref{fig:3}.}

\begin{figure}
\begin{center}
  \C{\includegraphics[scale=0.5]{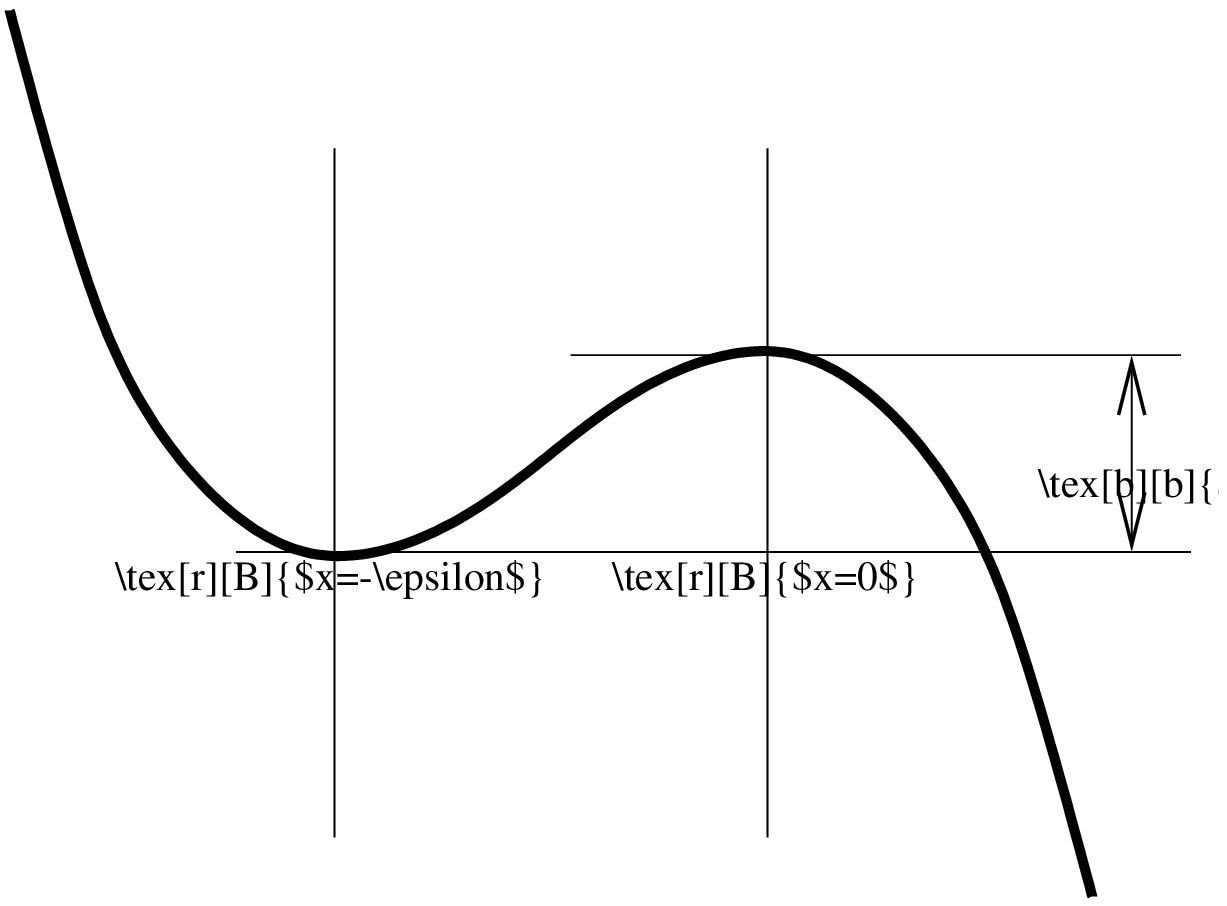} }
  \caption{The typical shape of the effective local potential
    $V(x)=-\epsilon x^2/2-x^3/3$ near
    $x=0$. Note that the depth of the potential \emph{and}  its width
  depend on $\epsilon $.}\label{fig:4}
  \end{center}
\end{figure}

To quantify this phenomenon, we assume that to lowest order, the
unstable manifold is moved by an amount $A\cdot(\eh -\epsilon)$
in the positive direction, \ie, $A>0$.
The potential along the $x$-axis is shown in \fref{fig:4}.


We next ask, for a fixed $\epsilon \in (0,\eh)$ and a
fixed $x\in [0,\epsilon ]$ how long the stochastic process \eref{e:1},
starting at $x$, needs to escape to the right (to $+\infty $). We will
neglect the $y$ coordinate in this estimate. As is well known, and for
example done in detail in Section~3 of
\cite{ETW_intermitt-noise_JPA1982}, see also \cite{freidlin-wentzell_stochdiffeq}, this time is given by
the Green's function of the differential operator $G$,
\eref{e:1}:
\begin{equ}\label{e:2}
  G= \frac{\sigma^2}{2}\frac{\d^2}{\dx ^2}+ (\epsilon x+x^2)\frac{\d}{\dx }~,
\end{equ}
with Dirichlet boundary condition at $x=+\infty $. The expected time $\tau(x)$
to escape from $x$ is then given by
\begin{equ}\label{e:double}
  \tau(x)= \frac{2}{\sigma^2}\int _x^\infty \d z\, e^{-h(z)}\, \int
  _{-\infty }^z \d w\, e^ {h(w)}~,
\end{equ}
with the potential $h=-V$ given by:
\begin{equ}
  h(z)= \frac{2}{\sigma^2} \left( \frac{\epsilon z^2}{2}+\frac{z^3}{3}\right)~.
\end{equ}
The integral \eref{e:double} can be estimated as in \cite{ETW_intermitt-noise_JPA1982}.
First one changes variables to $u=z+w$ and $v=z-w$ and finds
\begin{equ}\label{e:double2}
  \tau(-\infty)= \frac{2}{\sigma^2}\int _{-\infty}^\infty \d
  u\int_0^\infty  \d v\,
  \exp\left(-\frac{2}{\sigma^2}\bigl(\frac{1}{12}v^3+(\frac{\epsilon u
  }{2}+\frac{\epsilon u^2}{4}) v \bigr)\right)~.
\end{equ}
(Pushing the integration limit to $x=-\infty $ is justified by the
fact that anyway, most of the time is spent near $x=-\epsilon $.)
The $u$ integration can be done explicitly and leads to
\begin{equ}\label{e:double3}
  \tau(-\infty)= \frac{2}{\sigma^2}\int_0^\infty  \d v\,\bigl(\frac{\pi}{v}\bigr)^{1/2}
  \exp\left(-\frac{2}{\sigma^2}\bigl(\frac{v^3}{12}-\frac{\epsilon^2 v}{4}
   \bigr)\right)~.
\end{equ}
We rescale by $v=\epsilon w$ and thus find
\begin{equ}\label{e:double4}
  \tau(-\infty)= \frac{2\epsilon ^{1/2}}{\sigma^2}\int_0^\infty  \d w\,\bigl(\frac{\pi}{w}\bigr)^{1/2}
  \exp\left(-\frac{2\epsilon ^3}{\sigma^2}\bigl(\frac{w^3}{12}-\frac{ w}{4}
   \bigr)\right)~.
\end{equ}
Using the saddle-point approximation (the critical point is at $w=1$) this integral behaves, for large
$\epsilon^3 /\sigma^2$, and neglecting the prefactor in front of the exponential, as,
\begin{equ}
  \tau(-\infty ) \sim \exp\left( \frac{\epsilon ^3}{3\sigma^2 }\right)~.
\end{equ}

On the other hand, as illustrated in \fref{fig:3}, the probability to
reach a point $x<0$ is proportional to
$\exp(-\const(|x|+F)^2/\sigma^2)$,
where the constant $F$ depends on certain global aspects of the problem,
such as the length of the (almost) homoclinic loop. This just
estimates how much probability leaks to the ``wrong'', \ie, left side
of the unstable manifold of $(0,0)$. We will continue
the discussion by assuming all the constants to be 1. A rescaling of the
variables would eliminate an arbitrary constant anyway. In particular,
the average time to leave the trap (say, between $x=-\epsilon $ and
$x=0$) is then given approximately by
\begin{equ}\label{e:tescape}
\tau_{\rm escape}\sim \exp\left(\frac{\epsilon ^3}{3\sigma^2}-\frac{2}{\sigma^2}\bigl(A\cdot (\eh-\epsilon )\bigr)^2 \right)~.
\end{equ}
Here, we have used that, to lowest order, $D=A\cdot (\eh-\epsilon )$.

Consider now the polynomial in \eref{e:tescape}. It can be written as
\begin{equ}
  \frac{\epsilon^2 }{3\sigma^2}\bigl(\epsilon- 6 (A\cdot(\eh/\epsilon
  -1))^2 \bigr)~.
\end{equ}
For fixed $\eh$, this polynomial has
exactly one real root $\et=\et(A,\eh )$ which lies in
$(0,\eh)$. \emph{This is the point where the behavior will
  switch over.} It is the point which corresponds to $\ft$ in \fref{fig:forcevelocity}.
  
\section{Global topological aspects}
\label{sec:Global}

After having neglected the torus structure of the problem, we
reinstate it in the current section.
If we want to perform a global study of the system in the parameters
$\epsilon $ and $K_\perp$ we have to take into account that the phase space of
\eref{e:main1a} (or \eref{e:main1}) is a torus. Thus there can (and do) exist several
topologically different ways in which a homoclinic orbit can form.
They
can be indexed by two (non-negative) integers $\rho$ and $\psi$ which
count how many turns of $2\pi$ the variable $r$ resp.~$2\phi$ will
undergo as one moves from the fixed point $H_1$ to reach it again
through the homoclinic loop, and denote by $\W(\rho,\psi)$ the index of the orbit.

\begin{figure}[h]
  \centering
  \includegraphics[width=.6\columnwidth,angle=270]{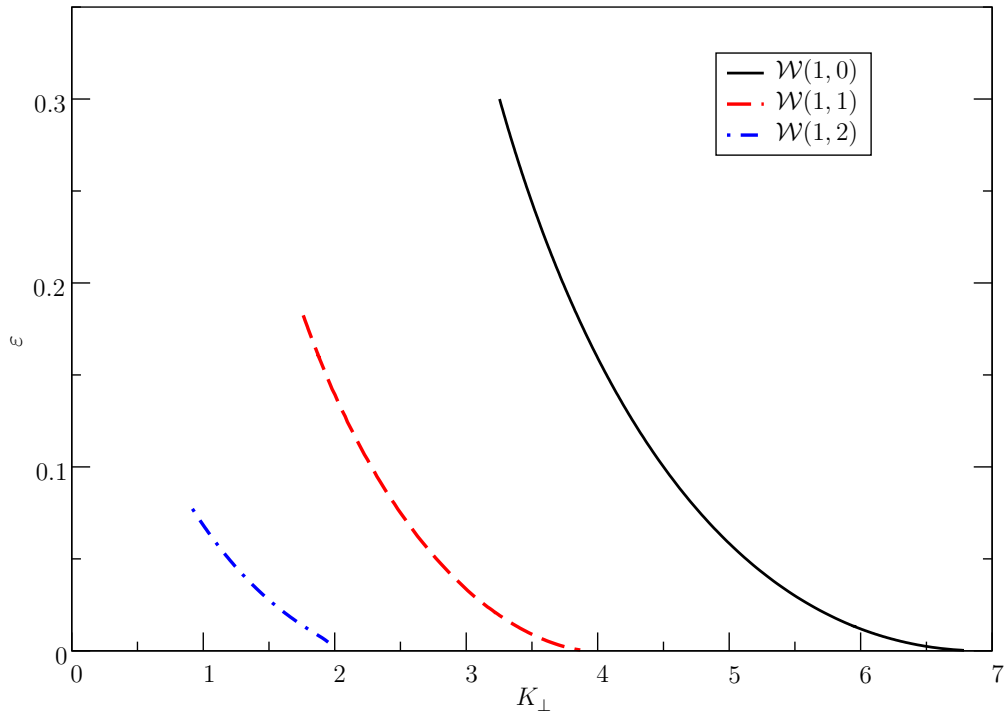}\quad
  \vspace*{-2mm}
  \caption{The locus of some homoclinic connections in the $\epsilon$,
    $K_\perp$ plane. Shown are 3 such curves with winding number in $r$
    equal to 1, and in $\phi$ equal to $0,1,2$. There are infinitely
    many such curves.}
  \vspace*{-1mm}
  \label{f:bifur}
\end{figure}

In the space of $\epsilon $ and $K_\perp$, the picture which emerges
numerically is shown in \fref{f:bifur}.
For each of the curves in \fref{f:bifur} we show one example
in \fref{fig:123}.
\def\ang{50}\def\sz{0.315\columnwidth}\def\kkk{\kern-10mm}
   \begin{figure}[h!]
    \C{\includegraphics[width=\sz,angle=\ang]{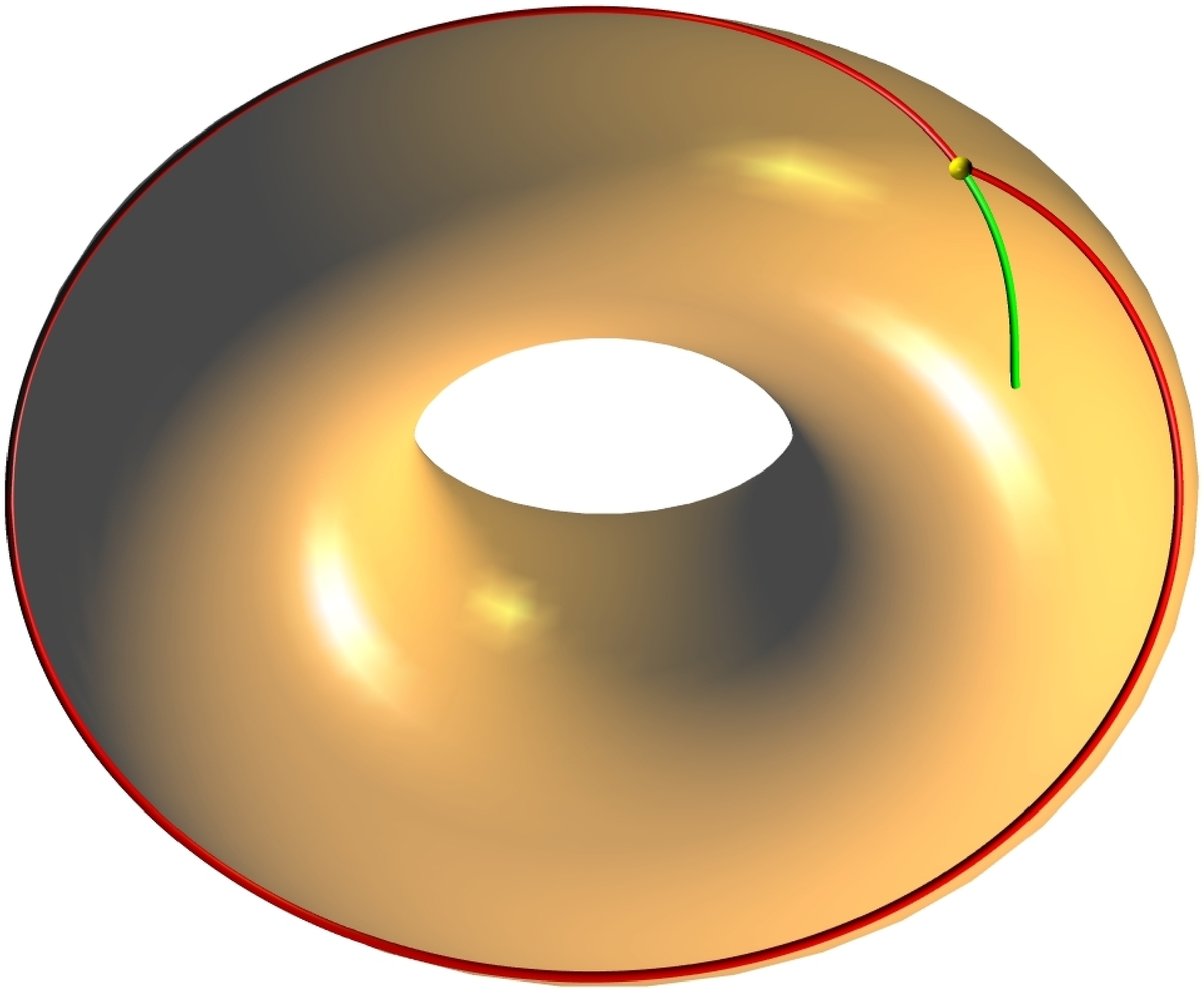}\kkk
\includegraphics[width=\sz,angle=\ang]{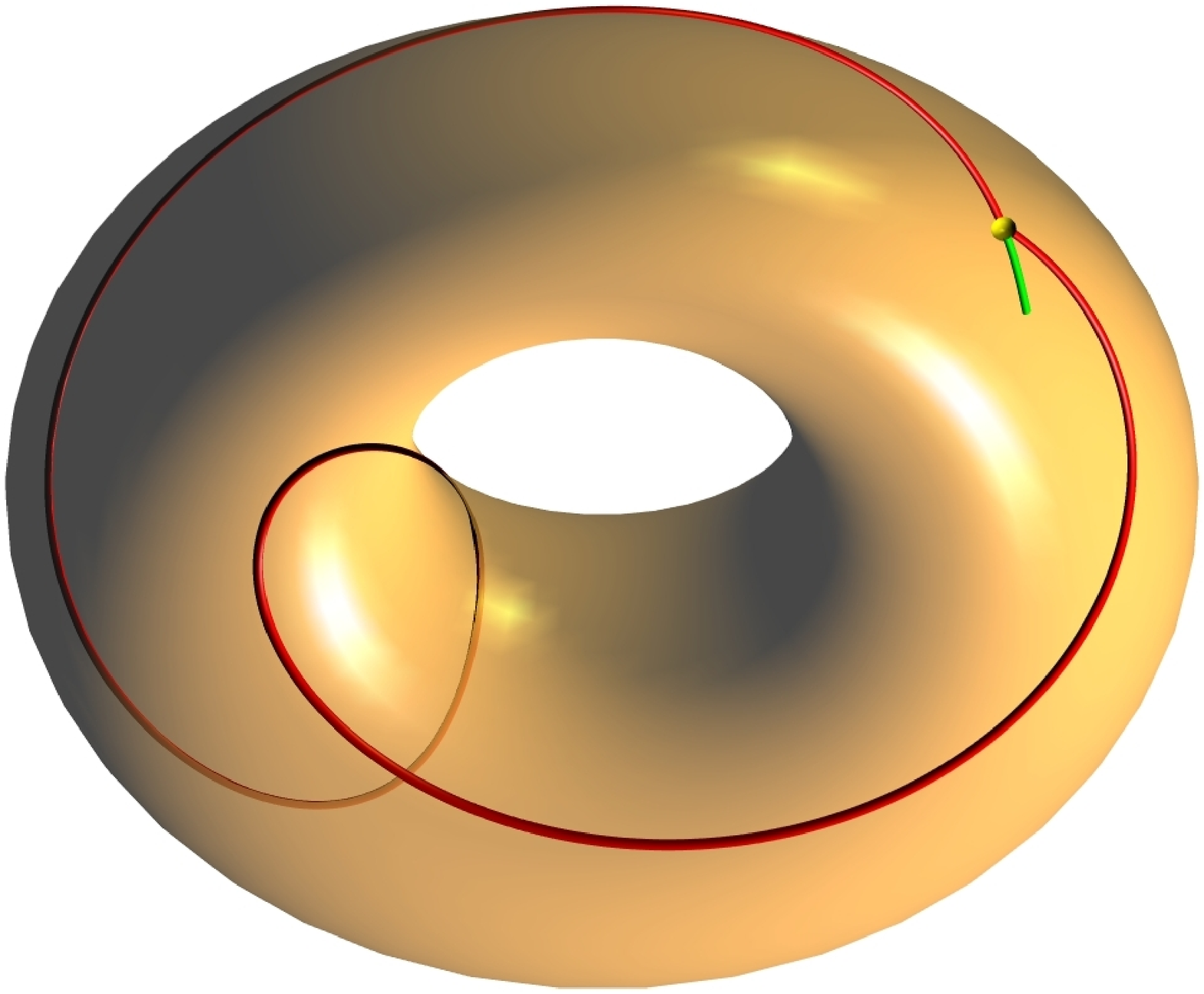}\kkk
\includegraphics[width=\sz,angle=\ang]{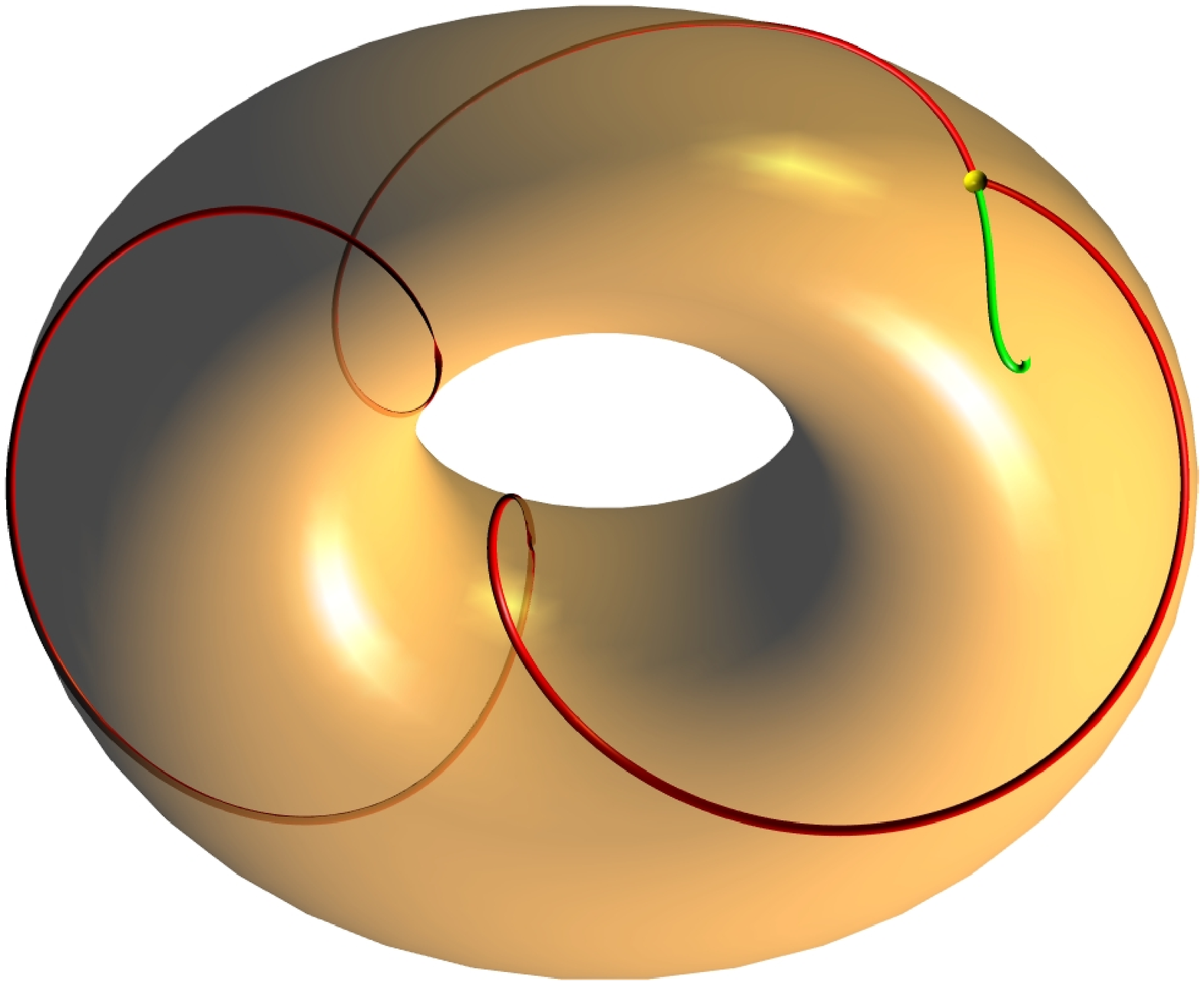}\quad\quad}
\vspace*{-10mm}
      \caption{From left to right: 3 homoclinic orbits, of type
        $\W(1,0)$, $\W(1,1)$, and $\W(1,2)$, respectively.}\label{fig:123}
    \end{figure}

\begin{figure}[h!]
      \C{\includegraphics[width=.7\columnwidth]{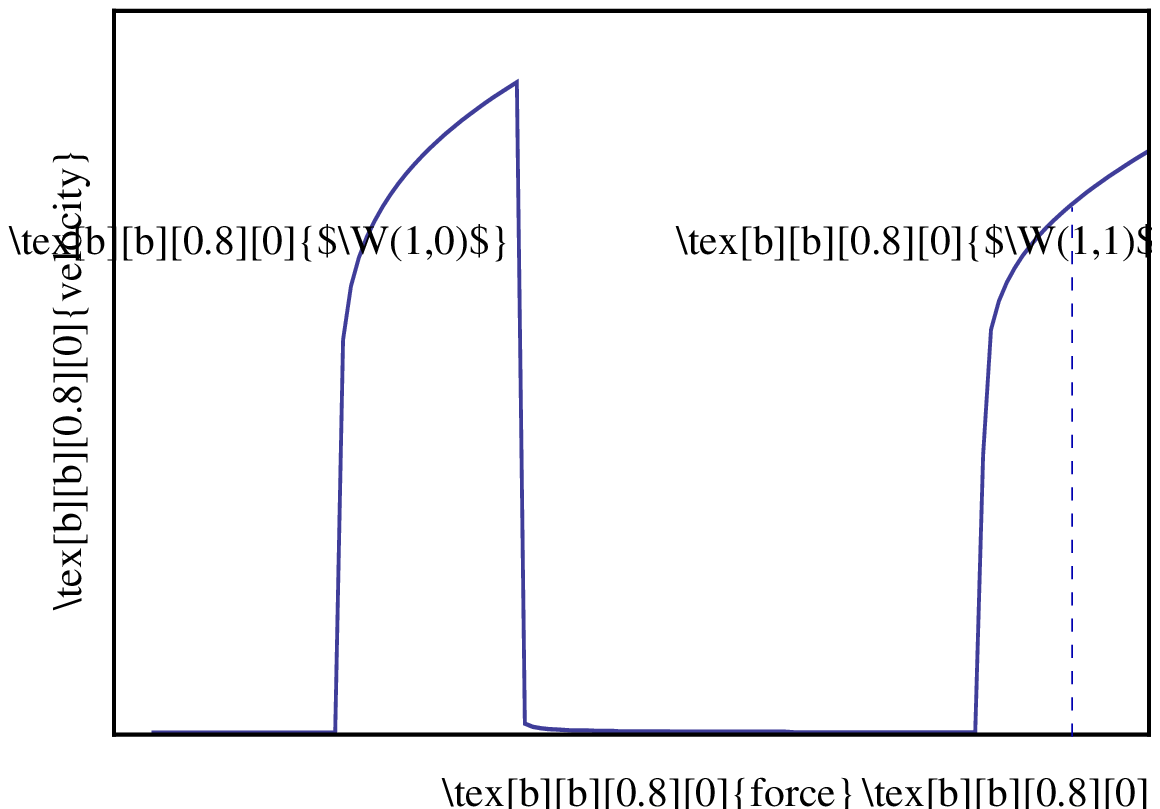}}
      \caption{A schematic illustration of the velocity as a function
        of force. First the topological type $\WW(1,0)$
      leads to a monotone increase of the velocity. But at some point,
      the winding number for $\phi$ changes, and the velocity drops to
      0. Several other winding numbers, not shown, could occur
      before the topological case $\WW(1,1)$ sets in.}\label{fig:bistability}
\end{figure}

It is now easy to explain  the non-monotonicity of the mean
velocity for \eref{e:main1}, as illustrated in \fref{fig:bistability}.
Fixing a $K_\perp$ (say $K_\perp=3.5$ in \fref{f:bifur}) and varying the  pulling
force $f=1-\epsilon $ for $\epsilon$ from $0$ to $0.3$ we first cross the $\W(1,1)$
curve and then the $\W(1,0)$ curve. This leads to
\fref{fig:bistability}. Note that, in accordance with the theory of
\sref{sec:General}, as a function of the noise (temperature), \emph{both} bumps are filled with
details which look as in
\fref{fig:forcevelocity}. In particular, there will be a special
bifurcation point of the form $\ft$ for both of them.

In \fref{fig:bistability}, we illustrate the cases
$\W(1,0) $ and $\W(1,1) $. The reader should notice
that for every pair $(\rho,\psi)$ which is realized, for fixed
$K_\perp$, there will be a window $\WW_{\rho,\psi}$ of values of $f$
around $\fh(\rho,\psi)$ which is like the case we discussed in detail.
\emph{Mutatis mutandis} our analysis will apply immediately to
all these cases, as soon as the general hypotheses (about the
transverse stability of the homoclinic orbit and the motion of the
return as a function of $f$) are satisfied.

The physical interpretation of the non-monotonous behavior of the
velocity is the following. We have seen previously (see
\sref{sec:results}) that $\phi$ ``helps'' $r$ to cross the
barriers between the local minima of the potential where
it lives. Doing so, $\phi$ oscillates in its own local minimum (this
corresponds to the first bump in \fref{fig:bistability}).  However,
for larger values of the force, $\phi$ will itself cross the barriers of its potential
and dissipate so much energy that it cannot help $r$ anymore (in the
phase space picture of \fref{fig:phase-portrait_f-above-fc}, this
corresponds to a collision between the attractive and repulsive limit
cycle). There is a whole regime of force where no limit cycle exists
(this is the flat region between the two bumps in \fref{fig:bistability}).
It is only for larger values of $f$ that a
stationary regime appears where both $\phi$ and $r$ can cooperate and display
non-zero mean velocity (this corresponds to the second bump in
\fref{fig:bistability}).
\medskip

\ack
This work was supported in part by the Swiss NSF under
MaNEP, Division II and partially by an ``Ideas''
advanced grant from the European Research Council.
We thank an unnamed referee for very useful
comments concerning the It\^o term, which we omitted in an earlier version.

\section*{References}
\bibliographystyle{JPE}
\bibliography{refs}

\end{document}